\documentclass[letterpaper]{article}

\usepackage{natbib,alifeconf}  
\usepackage{subcaption, float,hyperref}

\title{Emergence of more contagious COVID-19 variants from the coevolution of viruses and policy interventions}
\author{Aymeric Vi\'{e}$^{1,2}$\\
\mbox{}\\
$^1$Mathematical Institute, University of Oxford \\
$^2$Institute of New Economic Thinking, University of Oxford \\
vie@maths.ox.ac.uk}

\begin{document}
\maketitle

\begin{abstract}
At the end of 2020, policy responses to the SARS-CoV-2 outbreak have been shaken by the emergence of virus variants. The emergence of these more contagious, more severe, or even vaccine-resistant strains have challenged worldwide policy interventions. Anticipating the emergence of these mutations to plan ahead adequate policies, and understanding how human behaviors may affect the evolution of viruses by coevolution, are key challenges. In this article, we propose coevolution with genetic algorithms (GAs) as a credible approach to model this relationship, highlighting its implications, potential and challenges. We present a dual GA model in which both viruses aiming for survival and policy measures aiming at minimising infection rates in the population, competitively evolve. Simulation runs reproduce the emergence of more contagious variants, and identifies the evolution of policy responses as a determinant cause of this phenomenon. This coevolution opens new possibilities to visualise the impact of governments interventions not only on outbreak dynamics, but also on its evolution, to improve the efficacy of policies.
\end{abstract}

\section{Introduction}

The recent awareness on the emergence of variants has transformed the trajectory and impact of the SARS-CoV-2 outbreak. As early as June 2020, the initial COVID-19 strain identified in China was replaced as the dominant variant by the D614G mutation, found to have increased infectivity and transmission (\cite{WHO-variants}). On November 5 2020, a new strain of SARS-CoV-2 was reported in Denmark (\cite{WHO-mink}), linked with the mink industry, found to moderately decrease the sensitivity of the disease to neutralising antibodies. On 14 December 2020, the United Kingdom reported a new variant VOC 202012/01, with a remarkable number of 23 mutations, with unclear origin (\cite{sciencemag}). Early analyses have found that the variant has increased transmissibility, though no change in disease severity was identified (\cite{WHO-variants}). Quickly becoming dominant in Europe, this variant was held responsible for a significant increase in mortality, ICU occupation and infections across the country (\cite{wallace2021abrupt,iacobucci2021covid}). On 18 December 2020, the variant 501Y.V2 was detected in South Africa, after rapidly displacing other virus lineages in the region. 501Y.V2 was associated with a higher viral load, which may cause increased transmissibility (\cite{WHO-variants}), and found to undermine the efficacy of vaccines (\cite{mahase2021covid}).\\

Policy interventions against COVID-19 are changing objects as well, that can be seen as evolving towards greater efficacy through experimentation, learning and communication. As the virus mutates, so do the policy interventions, in a competitive coevolution process. The large search spaces involved, the focus on individual viruses, hosts ((\cite{guo2020evolutionary})), and inner focus on a genetic representation makes GAs adequate to model this competitive adaptation (\cite{holland1992genetic,lohn2002comparing,vie2021modelling}). It is common in coevolution that greater ecological pressures can increase the fitness of the populations involved (\cite{rosin1997new}). We are here examining this phenomenon with COVID-19 variants, evaluating the impact of policy interventions evolution over the evolution of the viruses. 

\section{Model}

Starting from initial conditions constituted by i) a population distribution of SARS-CoV-2 variants with identified genome sequences and traits and ii) a distribution of the current policy measures, we can simulate the evolution of viruses and policy actions as two coevolving GAs\footnote{All data and simulation code is available at \url{https://github.com/aymericvie/Covid19_coevolution}}. \par
We assume that viruses do not have any fitness function to maximise. Each virus is characterised by a reproduction rate, composed of a base rate and the sum of the impacts of its mutations. In initialisation, all viruses have no active mutations. Some mutations will increase the virus reproduction rate, some will be innocuous and others will decrease it. At each period of time, each individual virus genome replicates itself by infecting new individuals according to its reproduction rate, reduced by policies' effects. Each replication can trigger random binary mutations that (dis)activate specific mutations in the virus gene. \par
Policy interventions in the model are composed of 46 different non-pharmaceutical interventions, which effects on the virus reproduction rate have been identified (\cite{haug2020ranking}), from lockdowns to targeted closures. Policies are initialised with no active measure, and aim at minimising the virus reproduction rate. By fitness-proportionate selection, uniform crossover and random binary mutations, policies can activate different measures, increasing their efficacy. Policies reduce the effective reproduction rate of the viruses, controlling the outbreak. 


\section{Results}

\textbf{Under coevolution, virus adaptation towards more infectious variants is considerably faster than when the virus evolves against a static policy}. Although unguided by an objective, viruses evolve more efficiently facing a strong policy opposition (coevolution) than when the policies stay inactive (virus-only evolution). The average virus reproduction rate rises considerably more (up to 3.1) under coevolution than under virus-only evolution, in which this increase is low, and stays close to the natural reproduction rate of 2.63 (\cite{mahase2020covid}). Despite fewer hosts, selection in the virus population becomes more efficient under coevolution. 

\begin{figure}[H]
        \centering
        \begin{subfigure}[b]{0.415\textwidth}
            \centering
            \includegraphics[width=\textwidth]{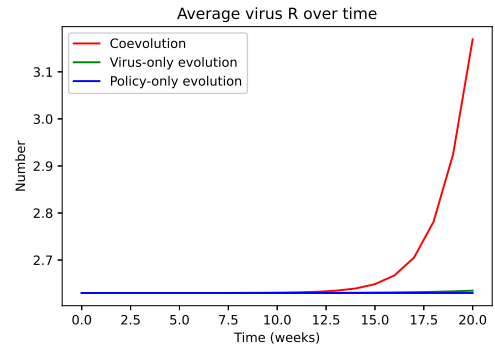}
            \caption[]
            {{\small Average reproduction rate of viruses over time}}   
            \label{virusR}
        \end{subfigure}
        \hfill
        \begin{subfigure}[b]{0.415\textwidth}  
            \centering 
            \includegraphics[width=\textwidth]{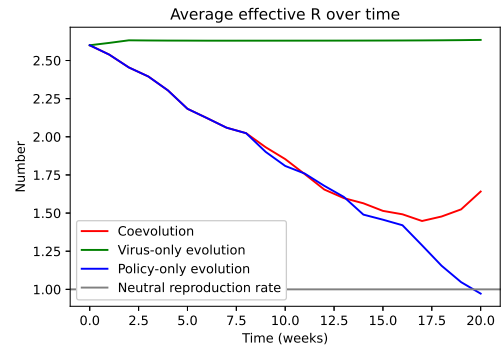}
            \caption[]
            {{\small Average effective reproduction rate over time}} 
            \label{effectiveR}
        \end{subfigure}
        \vskip\baselineskip
        \begin{subfigure}[b]{0.415\textwidth}   
            \centering 
            \includegraphics[width=\textwidth]{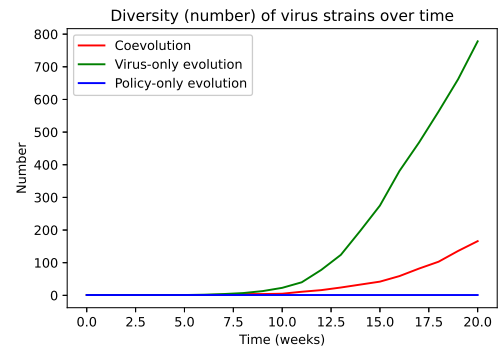}
            \caption[]
            {{\small Number of different virus strains over time}} 
            \label{diversity}
        \end{subfigure}
        \hfill
        \begin{subfigure}[b]{0.415\textwidth}   
            \centering 
            \includegraphics[width=\textwidth]{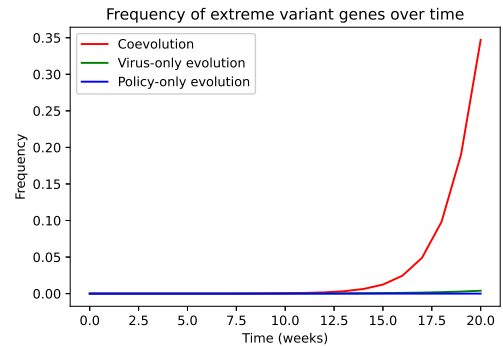}
            \caption[]
            {{\small Frequency of extreme variant genes over time}} 
            \label{virus_genome}
        \end{subfigure}
        \caption[Key results from the coevolution dual genetic algorithm]
        {\small Key results from the coevolution dual genetic algorithm} 
        \label{results}
    \end{figure}

\textbf{More contagious strains become dominant much faster in the virus population under coevolution}. 
Figure \ref{virus_genome} displays the frequency of viruses in the virus population containing the mutation gene granting the highest increase in reproduction rate. This fraction rises to 0.35 in the coevolution case, while staying considerably lower under virus-only evolution. This phenomenon is not driven by diversity and population size, but by a higher efficiency of evolution. Figure \ref{diversity} shows that the number of different variants in the population rises up to 800 under virus-only evolution, but only to 200 under coevolution. This difference is explained by the relatively large number of cases obtained under unconstrained virus-only evolution. As this extreme variant becomes dominant, the coevolution simulation run reproduced a variant-induced second wave of infections similar to the impact of VOC 202012/01 in the UK (Figure \ref{effectiveR}). Both relaxing measures for political or economical motives, and emergence of variants, can thus trigger multiple waves. 
    
\textbf{Seeing more infectious virus variants becoming dominant may signify that the policy measures are effective}. When policies are not evolving, more infectious variants are slower to become dominant in the population. Several countries today see contagious variants quickly attain such dominance. While this dynamic constitutes a new challenge, it can be seen as the sign that the current measures are putting stress on the virus: they are efficient in pushing weaker strains to extinction. Our future work with this model will strive to include vaccines as a policy measure, allow viruses to obtain a vaccine-resistant trait by mutations, and observe how the evolution of vaccine policies shapes the emergence of vaccine-resistant strains of SARS-CoV-2.  

\hyphenation{ap-proximated}

\footnotesize
\bibliographystyle{apalike}
\bibliography{example} 

\end{document}